\documentclass[twocolumn,aps,pra,showpacs,amsmath,amssymb,superscriptaddress]{revtex4}

\usepackage{graphicx}
\usepackage{dcolumn}
\usepackage{bm}
\usepackage{hyperref}
\usepackage{rotating}
\def\<{\langle}
\def\>{\rangle}

\graphicspath{{converted_graphics/}}

\begin{document}

\preprint{APS/123-QED}
\title{On the Appearance of Families of Efimov States in the Spinor Three-Body Problem}

\author{V. E. Colussi}
\address{Department of Physics, University of Colorado, Boulder, Colorado 80309-0440, USA}
\author{Chris H. Greene}
\address{ Department of Physics, Purdue University, West Lafayette, Indiana 47907-2036, USA}
\author{J. P. D'Incao}
\address{Department of Physics, University of Colorado, Boulder, Colorado 80309-0440, USA}
\address{ JILA, University of Colorado and NIST, Boulder, Colorado 80309-0440, USA}

\begin{abstract}
Few-body systems with access to multiple internal levels exhibit richness beyond that typically found in their single-level counterparts.  One example is that of Efimov states in strongly-correlated spinor three-body systems.  In [V. E. Colussi, C. H. Greene, and J. P. D'Incao, Phys. Rev. Lett. {\bf 113}, 045302 (2014)] this problem was analyzed for spinor condensates finding a complex level structure as in an early work [Bulgac and Efimov, Sov. J. Nucl. Phys. 22, 153 (1976)] in nuclear physics, and the impact of Efimov physics on the general form of the scattering observables was worked out.  In this paper we discuss the appearance of novel families of Efimov states in the spinor three-body problem.  

\end{abstract}

\pacs{67.85.Fg, 31.15.xj, 34.50.-s}

\maketitle

\section{\label{sec:level1}Introduction\protect\\}
The spinor three-body problem has received renewed interest recently due to the flourish of activity in the cold-atom community focused on the creation and manipulation of Bose-condensed gasses of alkali atoms with access to the full multiplet of hyperfine ground states (see Ref.~\cite{SpinorReview} and citations within.)  
These spinor condensates exhibit the rich interplay between superfluidity and magnetism, displaying interesting static and nonequilibrium many-body phenomenon from spin textures and spin domains to complex spin mixing dynamics \cite{SpinorReview}.  This scenario also displays richness on the few-body level, where recent work has focused on the strongly-correlated regime \cite{Colussi2014,Colussi2015}.  In spinor condensate experiments with alkali atoms to date, the interactions have typically been weak (with the exception of $^{85}$Rb \cite{Colussi2015,klausen2001PRA} 
and possibly also $^{133}$Cs and $^{7}$Li), 
and so the strongly-correlated regime remains largely unexplored.  In the strongly correlated regime, the $s$-wave scattering lengths associated with each two-body spin state exceeds the typical range of interatomic interactions, and Efimov physics becomes important \cite{efimov1970SJNP,braaten2006PR,wang2013AAMOP}. 
In the single-level case, this leads to the formation of Efimov states whose level spectrum has an exponential structure and whose presence strongly 
affects the scattering observables, characterized by the geometric scaling factor $e^{\pi/|s_0|}$, where $s_0$  ($\approx1.00624i$ for identical bosons) is a universal
constant that relates to the strength of the three-body interaction.  One of the major differences of the spinor three-body problem is the presence of multiple length scales in the problem associated with the $s$-wave scattering lengths for each of the two-body scattering channels. This not only modifies the energy spectrum but also increases the number 
of pathways three-body collisional processes like three-body recombination, atom-dimer relaxation and three-body spin-exchange,
can occur \cite{Colussi2014}.  In addition to that, the spin physics leads to the appearance of different values for the Efimov $s_0$ parameter,
leading to the formation of novel families of Efimov states \cite{Colussi2014,Colussi2015}.

In this paper we present a simple toy model which illuminates how these novel Efimov states arise in spinor systems.  We consider a model where 
each boson has access to two degenerate internal states.  By considering the full parameter space of scattering lengths and couplings between 
the two-body scattering channels, the mechanism by which the novel Efimov states arise is characterized.  
Our results echo the findings of Ref.~\cite{Colussi2014,Colussi2015,bulgacspinor}, although the toy model presented here is the 
first to probe the parameter space of scattering lengths and couplings in the case of degenerate internal levels.  

\section{\label{sec:level1}Background\protect\\}
On the one-body level the two internal states are labeled $|1\>$ and $|2\>$.  Pairs of atoms interact via a multichannel generalization of the Fermi pseudopotential \cite{efimovcontinuum,greensfunction,macek}, which (in a.u.) has the form:
\begin{equation}
\hat{v}(r)=\frac{4\pi\hat{A}}{m}\delta^3(\vec{r})\frac{\partial}{\partial r}r=
\frac{4\pi}{m}\left(\sum_{\sigma\sigma'}|\sigma\rangle A_{\sigma\sigma'}\langle\sigma'|\right)\delta^3(\vec{r})\frac{\partial}{\partial r}r,
\label{eq:pseudo}
\end{equation}
where $\delta^3(\vec{r})$ is the three-dimensional Dirac delta-function, 
$m$ is the atomic mass, and $\hat{A}$ is the scattering length matrix.  The matrix $\hat{A}$ is a multichannel generalization of the scattering length where the diagonal elements represent the background scattering lengths in each channel and the off-diagonal elements are the couplings.  The two-body basis is constructed by requiring them to be eigenkets of the symmetrization operator $\{|\sigma\>\}=\{|11\>,|12\>_S,|22\>\}$ \cite{product}.

We solve the three-body problem in the adiabatic hyperspherical representation, using the Green's function method developed in Refs~\cite{efimovcontinuum,greensfunction}.  The three-body system is characterized by the hyperradius $R$ which sets the overall size of the system and a set of five hyperangles collectively denoted $\Omega$ which describe the internal motion.  Treating $R$ as an adiabatic parameter, the three-body wave-function can be written as
\begin{equation}
\Psi(R,\Omega)=\sum_\nu F_\nu(R)\sum_\Sigma \Phi_\nu^\Sigma(R;\Omega)|\Sigma\>,\label{eq:threebodywfn}
\end{equation}
where $F(R)$ are the hyperradial wave functions, $\Phi(R;\Omega)$ the channel functions, and $\{|\Sigma\>\}=\{|111\>,|112\>_S,|211\>,|122\>,|221\>_S,|222\>\}$ the basis set labeling the internal levels for each three-body configuration \cite{product}.  
The task of solving the full three-body Schr\"odinger equation is reduced to solving a the fixed-$R$ hyperangular equation
\begin{multline} 
\sum_{\Sigma'}\left[\frac{\hat{\Lambda}^2(\Omega)+15/4}{2\mu R^2}\delta_{\Sigma\Sigma'}+\<\Sigma'|\hat{V}(R,\Omega)|\Sigma\>\right]\Phi_\nu^\Sigma(R;\Omega)\\
=U_\nu(R)\Phi_\nu^\Sigma(R;\Omega),\label{eq:schroad}
\end{multline}
where $\mu$ is the three-body reduced mass, $\hat{\Lambda}$ is the grand angular momentum operator \cite{avery}, and $\hat{V}$ is the sum of pairwise interactions.  The three-body potential $U_\nu(R)$ is obtained by solving Eq.~\ref{eq:schroad} for fixed $R$ values, where the problem reduces to solving a transcendental equation whose roots $s_\nu(R)$ enter as 
\begin{equation}
U_\nu(R)=\frac{s_\nu(R)^2-1/4}{2\mu R^2}.\label{eq:threebodypotential}
\end{equation}
In the single-level case for three identical bosons in the limit $R/a\rightarrow0$,  solving Eq.~\ref{eq:schroad} yields a single imaginary root $s_0\approx 1.00624i$, which when inserted in Eq.~\ref{eq:threebodypotential} gives an attractive $1/R^2$ potential which supports an infinite amount of bound trimers (Efimov states) characteristic of the Efimov effect.  For two identical and one dissimilar bosons $s_0\approx 0.41370i$.
In Ref.~\cite{Colussi2014,Colussi2015}, the first few $s_\nu(R)$ for spin-1,2, and 3 are tabulated over all relevant length scales.  The results demonstrate novel imaginary roots that lie in between the results from the analysis of single-level three-body systems as described in Ref.~\cite{bulgacspinor}.  In the next section we illustrate with a simple toy model how to understand the appearance of these novel roots.  
\section{\label{sec:level1}Toy Model\protect\\}
We will discuss first a toy model which explores the parameter space of the matrix $\hat{A}$, the toy model will then be generalized and a connection with the spinor three-body problem and the appearance of novel Efimov roots made.  In our two-level toy model $\hat{A}$ is a real-valued symmetric $3\times 3$ matrix
with eigenvalues $a_\alpha,a_\beta,$ and $a_\gamma$.  The full parameter space of $\hat{A}$ is large, but by considering a simple case the structure becomes more apparent.  Consider the scenario where $A_{1,3}=A_{2,3}=0$ with eigenvalues 
\begin{align}
a_{\alpha/\beta}&=\left(\frac{A_{1,1}+A_{2,2}}{2}\right)\pm\left[A_{1,2}^2+\left(\frac{A_{1,1}-A_{2,2}}{2}\right)^2\right]^{1/2}\nonumber\\
a_\gamma&=A_{3,3}\label{eq:egvalues}
\end{align}
and eigenvectors
\begin{align}
|\sigma_\alpha\>&=\cos\theta|11\>+\sin\theta |12\>_S\nonumber\\
|\sigma_\beta\>&=\sin\theta|11\>-\cos\theta |12\>_S\nonumber\\
|\sigma_\gamma\>&=|22\>\label{eq:evectors}
\end{align}
parameterized by the lone parameter
\begin{equation}
\tan\theta=\frac{2A_{1,2}}{\left[\left(A_{1,1}-A_{2,2}\right)+\left[4A_{1,2}^2+\left(A_{1,1}-A_{2,2}\right)^2\right]^{1/2}\right]}.
\end{equation}
By construction, variation of the angle $\theta\in\left[0,\pi/2\right ]$ ensures that the scattering lengths $a_\alpha,a_\beta,$ and $a_\gamma$ remain constant while the eigenvectors $|\sigma_\alpha\>$ and $|\sigma_\beta\>$ describe admixtures of the basis states $|11\>$ and $|12\>_S$.

In Figure 1, the results from solving Eq.~\ref{eq:schroad} for the first few roots $s_\nu(R)$ in the hyperradial region $a_\alpha\ll R\ll a_\beta$ are shown.  When $\theta=0$, there are two imaginary roots $s_0=s_1\approx 0.41370i$, associated with the three-body configurations $|221\>$, $|122\>$, and all other configurations obtained by label permutation.  When $\theta=\pi/2$, there is only one imaginary root $s_0\approx 1.00624i$ associated with the three-body configuration $|11|1\>$.  For intermediate values of $\theta$ the values $s_\nu$ are not 
assigned to any of single-level results. In fact, the $s_0$ and $s_1$ are associated with a three-body configuration which is a linear combination of the 
configurations of the extreme $\theta=0$ and $\pi/2$ cases and producing novel families of Efimov states.

\begin{figure}
\includegraphics[width=8cm,clip]{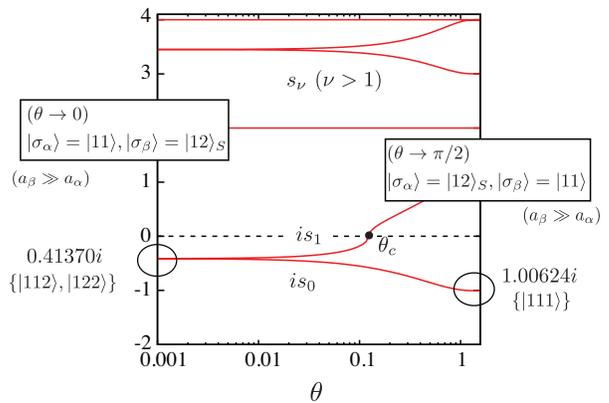}
\caption{The first few roots $s_\nu$ obtained from solving Eq.~\ref{eq:schroad} for the toy model in the hypperadial 
region $a_\alpha\ll R\ll a_\beta$.  The roots are plotted versus the angle $\theta$ which controls the amount of admixture in the system.} 
\end{figure}

The toy model is a result of a unitary transformation, $\hat{A}\rightarrow\hat{U}^\dagger\hat{A}\hat{U}$.  Equivalently, it is a result of a transformation (rotation) of 
the basis $\{|11\>,|12\>_S,|22\>\}\rightarrow\{|\sigma_\alpha\>,|\sigma_\beta\>,|22\>\}$.  What is then the class of unitary transformations which produce novel families of Efimov states?  The eigenvector of $\hat{A}$ attached to the resonant channel must mix two-body product states under the transformation, which cannot arise from a unitary transformation on the one-body level.

There is a simple connection between these results and those from Ref.~\cite{Colussi2014,Colussi2015} for spin-1, 2, and 3.  When the internal levels are composed of the hyperfine spin multiplet there is the additional requirement that the two-body eigenstates be simultaneous eigenstates of both the symmetrization operator and spin operator $\hat{F}^2$.  The unitary matrix that diagonalizes the matrix $\hat{F}^2$ carries the product basis $|m_{f_1}\>\otimes |m_{f_2}\>$ into the $|F_{2\mathrm B},m_{F2\mathrm b}\>$ basis, and the elements of this matrix are Clebsch-Gordan coefficients, which is also equivalent to a unitary transformation of the matrix $\hat{A}$.  In the toy model, the admixture was set by the angle $\theta$ and treated as tunable for exploring the full parameter space.  In the spinor case, this admixture is set by the values of the Clebsch-Gordan coefficients.  As in the toy model, this admixture produces roots $s_\nu$ which are associated with three-body spin configurations that can be composed of a linear combination of product state spin functions and are capable of producing novel families of Efimov states.  

\section{Acknowledgements}
This work was supported by the U.S. National Science
Foundation, grant numbers PHY-1125844, PHY-1307380
and PHY-1306905.


\begin{thebibliography}{}
\bibitem{product}  The ket $|11\>$ is shorthand for the tensor product $|1\>\otimes|1\>$.  The `$S$' subscript in $|12\>_S$ is shorthand for the symmetrization $\left(|12\>+|21\>\right)/\sqrt{2}$.  Similarly, for the three-body ket $|112\>_S=\left(|112\>+|121\>\right)/\sqrt{2}$.
\end{thebibliography}
\end{document}